\documentclass[10pt,pra,showkeys,showpacs,groupedaddress,twocolumn]{revtex4-1}
\usepackage{graphicx}
\usepackage{subfigure}
\usepackage{ulem}
\usepackage{amsmath}
\usepackage{amssymb}
\usepackage{wrapfig}
\usepackage[utf8]{inputenc}  
\usepackage{color} 

\begin{document}

\title{Electromagnetically-induced transparency, absorption, and microwave field sensing in a Rb vapor cell with a three-color all-infrared laser system}
\author{N.~Thaicharoen$^{1, \dag}$}
\thanks{Corresponding author: nithi@umich.edu}
\author{K.~R.~Moore$^{1, \dag\dag}$}
\author{D.~A.~Anderson$^{2}$}
\author{R. C. Powel$^{1, \dag\dag\dag}$}
\author{E. Peterson$^{1, \dag\dag\dag\dag}$}
\author{G.~Raithel$^{1, 2}$}
\affiliation{$^1$Department of Physics, University of Michigan, Ann Arbor, Michigan 48109, USA}
\affiliation{$^2$Rydberg Technologies Inc., Ann Arbor, Michigan 48104, USA}

\date{\today}

\begin{abstract}
A comprehensive study of three-photon electromagnetically-induced transparency (EIT) and absorption (EIA) on the rubidium cascade $5S_{1/2} \rightarrow 5P_{3/2}$ (laser wavelength 780~nm), $5P_{3/2} \rightarrow 5D_{5/2}$ (776~nm), and $5D_{5/2}\rightarrow 28F_{7/2}$ (1260~nm) is performed. The 780-nm probe and 776-nm dressing beams are counter-aligned through a Rb room-temperature
vapor cell, and the 1260-nm coupler beam is co- or counter-aligned with the probe beam. Several cases of EIT and EIA, measured over a range of detunings of the 776-nm beam, are studied. The observed phenomena are modeled by numerically solving the Lindblad equation, and the results are interpreted in terms of the probe-beam absorption behavior of velocity- and detuning-dependent dressed states. To explore the utility of three-photon Rydberg EIA/EIT for microwave electric-field diagnostics, a sub-THz field generated by a signal source and a frequency quadrupler is applied to the Rb cell. The 100.633-GHz field resonantly drives the $28F_{7/2}\leftrightarrow29D_{5/2}$ transition and causes Autler-Townes splittings in the Rydberg EIA/EIT spectra, which are measured and employed to characterize the performance of the microwave quadrupler.
\end{abstract}


\maketitle

\section{Introduction}

Rydberg levels of atoms and molecules are characterized by a tenuously-bound valence electron whose marginal atomic binding leads to high susceptibilities to external fields and other perturbations~\cite{Gallagher}. Electric-dipole transitions between Rydberg states are in the microwave and sub-THz range, with electric-dipole matrix elements scaling as the square of the principal quantum number $n$. Hence, Rydberg atoms exhibit a strong response to applied DC and radio-frequency (RF) electric fields. Based on these properties, Rydberg atoms are now being used and proposed widely in atomic measurement standards for electric fields~\citep{sedlacek_microwave_2012, gordon_millimeter_2014, holloway_sub-wavelength_2014, Holloway2014, miller_radio-frequency-modulated_2016, horsley_frequency-tunable_2016, kwak_microwave-induced_2016, AndersonCont2017} and in Rydberg-atom-based communications~\cite{AndersonRadioArxiv.2018, Debpub.2018, Meyerpub.2018}, with
cell-internal structures providing enhanced sensitivity~\cite{Anderson2018b}.
In the employed method of Rydberg-EIT~\cite{Mohapatra2007, Mauger2007}, a coupling laser resonantly couples a low-lying intermediate level, $\vert e \rangle$, to one or more Rydberg levels, $\vert r_i \rangle$, thereby inducing electromagnetically induced transparency (EIT~\citep{boller_observation_1991, fleischhauer_electromagnetically_2005}) for a probe beam that measures absorption on the transition between $\vert g \rangle$ and $\vert e \rangle$.
Rydberg interactions in cold-atom Rydberg-EIT have been measured~\cite{Weatherill2008,Pritchard2010} and
theoretically investigated~\cite{Petrosyan2011,Ates2011}. In the present study in room-temperature vapor cells, the observed Rydberg-EIT spectra serve as an optical probe for the energy levels of the Rydberg states $\vert r_i \rangle$, as well as for their energy-level shifts in applied DC and RF electric fields. Owing to the simplicity of vapor-cell spectroscopy, Rydberg-EIT-based atomic field measurement in room-temperature cesium and rubidium vapor cells is particularly attractive for novel metrology approaches that harness the quantum-mechanical properties of atoms.
Efforts are underway to develop the method into atom-based, calibration-free, sensitive field measurement and receiver instrumentation.

Since the matrix elements for optical Rydberg-atom excitation, $\langle r_i \vert {\hat{\bf{r}}} \vert e \rangle$, are quite small, two-color Rydberg-EIT as described above often requires expensive commercial lasers for the coupling transition.
For instance, in Rb and Cs Rydberg-EIT one typically requires a coupling laser at respective wavelengths of 480~nm and 510~nm, $\sim$1~MHz linewidth, tens of mW of power, and good spatial mode quality. There is an interest in replacing the two-color EIT with schemes involving three low-power infrared lasers instead~\citep{johnson_absolute_2010, carr_three-photon_2012, shaffer2018}.

In our paper, we compare experimental and theoretical results in three-color EIT and EIA (electromagnetically-induced absorption) in a Rb vapor cell. Several schemes with different relative propagation directions of the three optical beams and in intermediate-transition detuning values are studied, and regimes suitable for three-photon EIT and EIA spectroscopy of Rydberg energy levels are identified. The results are discussed in context with related works. We further demonstrate the utility of the setup for characterizing a commercial sub-THz frequency quadrupling system. Numerical solutions of the Lindblad equation and an analytical dressed-state approach are employed to model and interpret the data.

\section{Experimental Setup}

\begin{figure}[htb]
\centering
\includegraphics[width=\linewidth]{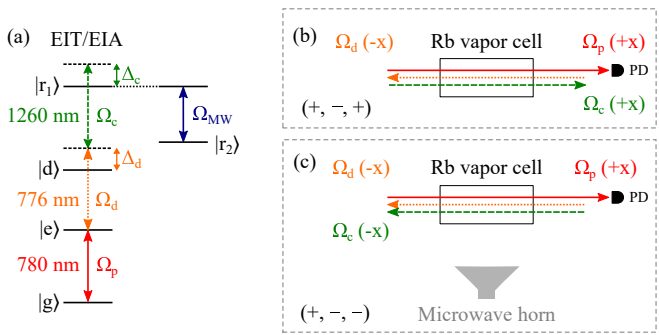}
\caption{(a) Energy-level diagram for 3-photon EIT experiment. (b-c) Experimental setup. The 780-nm beam counter-propagates with the 776-nm beam. The 1260-nm beam propagates either (b) in the same direction as the 780-nm beam or (c) in the opposite direction. The propagation-direction configurations for the 780-nm, 776-nm and 1260-nm beams are denoted ($+, -, +$) and ($+, -, -$), respectively.}
\label{fig:1}
\end{figure}
The energy-level diagram of the experiment is shown in Fig.~\ref{fig:1}~(a). Following the terminology in~\cite{carr_three-photon_2012}, the three transitions from the ground into the Rydberg levels are referred to as the probe, the dressing and the coupler transitions. All laser sources are home-built external-cavity diode lasers. The 780-nm probe laser is locked to the $\vert g \rangle \left(5S_{1/2}, F=3\right) $ $ \rightarrow \vert e \rangle  \left(5P_{3/2}, F'=4\right)$ transition at zero atomic velocity. Part of the locked 780-nm beam is sent into another rubidium reference vapor cell and counter-propagated with a small portion of the 776-nm laser to form a saturated spectroscopic signal to lock the 776-nm dressing laser, which is set at selected detunings $\Delta_d$ from the $\vert e \rangle  \rightarrow \vert d \rangle  \left(5D_{5/2}\right) $ transition at zero velocity. (The $5D_{5/2}$ hyperfine levels were not resolved in the present setup). The 1260-nm coupling laser is scanned through the $\vert d \rangle \rightarrow \vert r_1 \rangle \left( 28F_{7/2} \right)$ transition. The offset frequency of the 1260-nm laser, $\Delta_c$,
is calibrated by recording the transmission of a small fraction of the 1260-nm beam through a Fabry–P\'{e}rot cavity with a free spectral range of 375~MHz. The power of the transmitted probe beam is measured with a photodiode as a function of $\Delta_c$.

The 780-nm probe beam has a power of $P \lesssim 10~\mu$W and a Gaussian beam-waist parameter $w_{0}$ of $\gtrsim 80~\mu$m, corresponding to a Rabi frequency $\Omega_p \lesssim 2 \pi \times 22$~MHz. For the 776-nm dressing beam, $P \lesssim 3$~mW, $w_{0} \gtrsim 140~\mu$m, and $\Omega_d \lesssim 2 \pi \times 67$~MHz, and for the 1260-nm coupling beam $P \lesssim 7$~mW, $w_{0} \gtrsim 75~\mu$m, and $\Omega_c \lesssim 2 \pi \times 27$~MHz. The listed Rabi frequencies are averages over the relevant magnetic transitions and are calculated for the respective beam centers. The actual effective Rabi frequencies are considerably lower due to averaging over the near-Gaussian spatial beam profiles, possible imperfections in the beam overlaps, and possible beam-size increases due to lensing in the walls of the vapor cell.

The three laser beams must be carefully aligned and overlapped within the $L=7.5$-cm long Rb vapor cell. The first alignment step is to establish two-photon EIT by counter-propagating the 780-nm probe with the 776-nm dressing beam. This couples the lower three levels, $\vert g \rangle \leftrightarrow \vert e \rangle \leftrightarrow \vert d \rangle$. We optimize the $5D_{5/2}$ EIT signal by fine-adjusting the overlap between the probe and dressing beams and adjusting the power of the 780-nm probe beam.
Then, we apply the 1260-nm coupler beam either in the same direction as the 780-nm probe beam [($+, -, +$) configuration, see Fig.~\ref{fig:1} (b)] or in the direction opposite to the 780-nm beam [($+, -, -$) configuration, see Fig.~\ref{fig:1} (c)]. We fix the frequency of the 780-nm probe and the 776-nm dressing beams while scanning $\Delta_c$. The EIA or EIT signals are observed on top of the $5D_{5/2}$-EIT background.

To show the utility of three-photon EIT and EIA in probing microwave and sub-THz electric fields, we have calibrated the electric-field strength of a 100-GHz transmission system. A microwave source supplies a 25-GHz signal to an active quadrupler, which feeds 100-GHz radiation to a standard-gain horn. The three-photon EIT/EIA field probe is placed in the far field of the horn, as indicated in Fig.~\ref{fig:1}~(c). We test EIT- and EIA-schemes to calibrate the 100-GHz electric field against the 25-GHz power the signal generator supplies to the quadrupler.

\section{Numerical model}
\label{sec:model}

\subsection{Model outline}
\label{subsec:modelA}

The system is modeled with the five-level system shown in Fig.~\ref{fig:1}~(a).
We numerically solve the Lindblad equation of the system and obtain steady-state solutions of the
density-matrix, $\hat{\rho}$. We ignore magnetic substructure, other than including $m$-averaged angular matrix elements in the calculation of the Rabi frequencies. We assume a closed decay scheme in which $\vert e \rangle$ decays at a rate of $\Gamma_e = 2 \pi \times 6$~MHz into $\vert g \rangle$,  $\vert d \rangle$ at a rate of $\Gamma_d=2 \pi \times 0.66$~MHz into  $\vert e \rangle$, $\vert r_1 \rangle$ at a rate of $\Gamma_{r1}= 2 \pi \times 10$~kHz into $\vert d \rangle$, and $29D_{5/2}$ ($\vert r_2 \rangle$) at a rate of $\Gamma_{r2} = 2 \pi \times 10$~kHz into $\vert e \rangle$.
We neglect the (minor) decay of $5D_{5/2}$ ($\vert d \rangle$) into $6P_{3/2}$ and the decays of the Rydberg levels out of the five-level system. The system has four coherent-drive fields. The probe is linearly polarized in the horizontal direction, while all other fields are polarized vertically. The Rabi frequencies at the beam centers are calculated from the beam parameters provided above, the known radial electric-dipole matrix elements for the various transitions, and an average of the angular matrix elements over the relevant magnetic transitions.

We attribute the good agreement between calculated and measured results in Secs.~\ref{sec:EIT} and~\ref{sec:onephoton} to the absence of low-lying population-trapping metastable states. Also, the short ($\mu$s-long) atom-field interaction times in the room-temperature vapor cell negate significant Rydberg-level decay out of the closed five-level system assumed in the calculation. The exact values of the Rydberg-level decay rates used in the model have no significant effects on the results. Also, while additional level dephasing is included as an option in the model, this was not needed in order to reproduce the experimental data to within the experimental confidence levels. Another reason for the success of the five-level model is the absence of significant magnetic fields. Fields exceeding $\sim$1~Gauss would introduce Zeeman splittings and complex optical-pumping dynamics~\cite{Zhang2018} that cannot be captured in a five-level model.

\subsection{Formalism}
\label{subsec:modelB}

For a given set of probe, dressing, coupler and (optional) RF Rabi frequencies, $\Omega_p$, $\Omega_d$, $\Omega_{c}$
and $\Omega_{RF}$, and respective zero-velocity atom-field-detunings, $\Delta_p$, $\Delta_d$, $\Delta_c$ and $\Delta_{RF}$, we obtain the steady-state solution of the Lindblad equation in a four-color field picture.
Since our probe Rabi frequencies are larger than the $5P_{3/2}$ ($\vert e \rangle$) decay rate, we do not make a weak-probe approximation.
The atom-field detunings are defined as field frequencies minus atomic-transition frequencies. Accounting for the Doppler effect, the detunings $\Delta_{i,a}$ with $i=p, \, d, \, c, \, RF$, are, in the four-color field picture and in the frame of reference that is co-moving with the atom,
\begin{eqnarray}
\Delta_{p,a} & = & \Delta_p - k_p v \nonumber \\
\Delta_{d,a} & = & \Delta_d + k_d v \nonumber \\
\Delta_{c,a} & = & \Delta_c \pm k_c v \nonumber \\
\Delta_{RF,a} & = & \Delta_{RF} - k_{RF \, \parallel} v \quad,
\end{eqnarray}
where $v$ denotes the atom velocity along the probe-beam direction, the wavenumbers $k_i$ are defined as positives, and the term $\pm k_c$ corresponds with the ($+, -, \mp$) configurations, respectively. The wave-vector component of the RF field in the direction of the laser beams, $k_{RF \, \parallel}$, is so small that it can be neglected. In the Lindblad equation
\begin{equation}
\dot{\hat{\rho}} = \frac{i}{\hbar} [\hat{\rho},\hat{H} ] + L(\hat{\rho}) \quad ,
\end{equation}
the Hamiltonian matrix is
\begin{equation}
H = \hbar
\left(
  \begin{array}{ccccc}
 -\Delta_1             & \Omega_{p}/2          & 0             &            0 &    0 \\
 \Omega_{p}/2   & - \Delta_2            & \Omega_{d}/2  &            0 &    0 \\
  0             & \Omega_{d}/2          & -\Delta_3     & \Omega_{c}/2 &    0 \\
  0             & 0                     & \Omega_{c}/2  & -\Delta_4    &    \Omega_{RF}/2\\
  0             & 0                     & 0             &\Omega_{RF}/2 &    -\Delta_5
  \end{array}
\right) \quad .
\end{equation}
There, the Hamiltonian is expressed in the dressed-state basis $\{ \vert 1 \rangle, ..., \vert 5 \rangle \}$ that corresponds with the bare atomic states $\{ \vert g \rangle,  \vert e \rangle, \vert d \rangle, \vert r1 \rangle, \vert r2 \rangle \}$, in that order. The field-free dressed-state energies are
$\Delta_1=0$, $\Delta_2=\Delta_1 + \Delta_{p,a}$, $\Delta_3=\Delta_2+\Delta_{d,a}$, $\Delta_4=\Delta_3+\Delta_{c,a}$, and
$\Delta_5=\Delta_4 - \Delta_{RF,a}$. The fifth state is not used when the RF field is off.
For the Lindblad operator $L(\hat{\rho})$ we use the level decay rates given in Sec.~\ref{subsec:modelA}, with no additional level dephasing terms,
\begin{widetext}
\begin{equation}
L(\hat{\rho}) =
\left(
  \begin{array}{ccccc}
 \Gamma_e \rho_{22}
 & -\frac{1}{2} \Gamma_e \rho_{12}
 & -\frac{1}{2} \Gamma_d \rho_{13}
 & -\frac{1}{2} \Gamma_{r1} \rho_{14}
 & -\frac{1}{2} \Gamma_{r2} \rho_{15}  \\
   -\frac{1}{2} \Gamma_e \rho_{21}
 & -\Gamma_e \rho_{22}  + \Gamma_d \rho_{33} + \Gamma_{r2} \rho_{55}
 & -\frac{1}{2} (\Gamma_e + \Gamma_d   ) \rho_{23}
 & -\frac{1}{2} (\Gamma_e + \Gamma_{r1}) \rho_{24}
 & -\frac{1}{2} (\Gamma_e + \Gamma_{r2}) \rho_{25} \\
   -\frac{1}{2} \Gamma_d \rho_{31}
 & -\frac{1}{2} (\Gamma_e + \Gamma_d   ) \rho_{32}
 & -\Gamma_d \rho_{33}  + \Gamma_{r1} \rho_{44}
 & -\frac{1}{2} (\Gamma_d + \Gamma_{r1}) \rho_{34}
 & -\frac{1}{2} (\Gamma_d + \Gamma_{r2}) \rho_{35} \\
   -\frac{1}{2} \Gamma_{r1} \rho_{41}
 & -\frac{1}{2} (\Gamma_e + \Gamma_{r1}) \rho_{42}
 & -\frac{1}{2} (\Gamma_d + \Gamma_{r1}) \rho_{43}
 & -\Gamma_{r1} \rho_{44}
 & -\frac{1}{2} (\Gamma_{r1}+\Gamma_{r2}) \rho_{45} \\
   -\frac{1}{2} \Gamma_{r2} \rho_{51}
 & -\frac{1}{2} (\Gamma_e + \Gamma_{r2}) \rho_{52}
 & -\frac{1}{2} (\Gamma_d + \Gamma_{r2}) \rho_{53}
 & -\frac{1}{2} (\Gamma_{r1}+\Gamma_{r2}) \rho_{54}
 & - \Gamma_{r2} \rho_{55}
  \end{array}
\right) \quad.
\end{equation}
\end{widetext}
The steady-state solution
for $\hat{\rho}$ yields the coherence $\rho_{12}$ as a function of atom velocity $v$ and all field-strength, field-detuning and decay parameters. The absorption coefficient and the refractive index of the atomic vapor for the probe beam then follow
\begin{eqnarray}
 \alpha & = & \frac{\omega_p}{c} \frac{2 n_V d_{eg}}{\epsilon_0 E_P} \int P(v) {\rm{Im}} (\rho_{12}) dv \nonumber \\
  (n-1) & = & \frac{1}{2} \frac{2 n_V d_{eg}}{\epsilon_0 E_P}        \int P(v){\rm{Re}} (\rho_{12}) dv \quad.
\end{eqnarray}
Here, $\omega_p = k_p c$, $n_V$ denotes the atom volume density, $d_{eg}$ the probe electric-dipole matrix element, $E_P$ the probe-laser electric-field amplitude, and $P(v)$ the normalized one-dimensional Maxwell velocity distribution in the room-temperature vapor cell. In the $n_V$-value we account for the natural abundance of $^{85}$Rb in our cell (72$\%$) and the statistical weight of $^{85}$Rb $F=3$ (58.3$\%$). For the probe electric-dipole matrix element
averaged over the magnetic transitions we use $d_{eg} = 1.9$~ea$_0$.

\section{Three-photon EIA and EIT}
\label{sec:EIT}

\subsection{($+, -, -$) configuration}

\subsubsection{Measurement and simulation results}

The first objective of the work is to identify beam-directions, Rabi frequencies and detunings that yield EIT and EIA signatures suitable to measure Rydberg energy level positions and shifts. We find several regimes of robust EIA and EIT for the beam-propagation configurations $(+, -, \mp)$ defined in Fig.~\ref{fig:1}.

The $(+, -, -)$ configuration has been studied in~\cite{carr_three-photon_2012} for a case in cesium and $\Delta_d=0$. The Rb case studied here differs from~\cite{carr_three-photon_2012} in that the differential probe-dressing Doppler shift, $(k_d - k_p) v$, is near zero for a wide range of velocities within the Maxwell velocity distribution, because the probe and dressing wavelengths are near-identical in the Rb case studied experimentally in this work. This leads to a stronger EIA signal.  We further also explore the behavior at non-zero $\Delta_d$.

\begin{figure}[htb]
\centering
\includegraphics[width=0.48\textwidth]{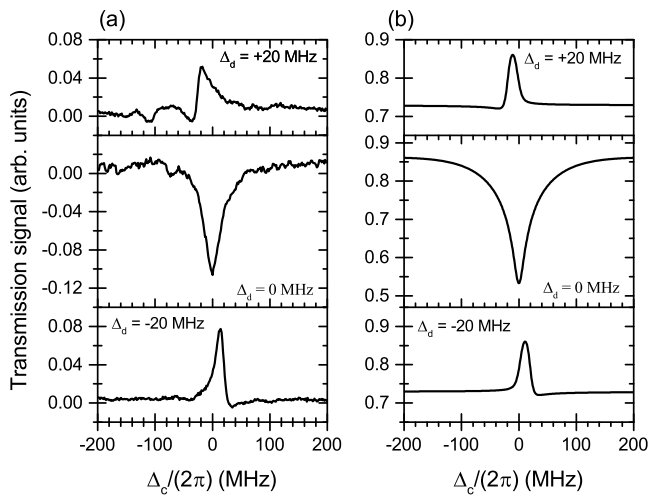}
\caption{(a) Change in experimental probe transmission vs coupler-laser detuning when the lasers are in ($+, -, -$) configuration. The three panels are on the same (arbitrary) scale. Strong EIA is observed when $\Delta_d = 0$, and EIT when $\Delta_d = \pm 2 \pi \times 20$~ MHz. (b) Probe transmissions calculated with the model in Sec.~\ref{sec:model} for cell temperature of 300~K, cell length $L=7.5$~cm and with $\Omega_p =  2 \pi \times 10$~MHz, $\Omega_d =  2 \pi \times 25$~MHz, and $\Omega_c = 2 \pi \times 18$~MHz.}
\label{fig:2}
\end{figure}

Fig.~\ref{fig:2}~(a) shows experimental results for the ($+, -, -$) configuration and $\Delta_d/(2 \pi) = -20, \, 0,$ and $+20$~MHz. The sub-THz field is turned off in this study. The examples shown in the figure illustrate our observation of strong EIA when $\Delta_d$ is close to zero and EIT when $\Delta_d$ is $\gtrsim 2 \pi \times 5$~MHz. Simulation results for this configuration are shown in Fig.~\ref{fig:2}~(b). The simulated results show $\exp(-\alpha L)$, with cell length $L=7.5$~cm and the absorption coefficient $\alpha$ computed for cell temperature 300~K using Eqs.~(1-5). The Rabi frequencies in the simulation were $\Omega_{p}=2 \pi \times 10$~MHz, $\Omega_{d}=2 \pi \times 25$~MHz, and $\Omega_{c}=2 \pi \times 18$~MHz; these values lead to good agreement between simulated and experimental data. The EIA and EIT line shapes, line widths and signal depths agree well between the experimental and simulated data.
(The experimental data show change in transmission on the same (arbitrary) scale for the different cases of $\Delta_d$.)
It is further seen, both in the experimental and the simulated data, that the EIT linewidth at non-zero $\Delta_{d}$ is smaller than the width of the EIA dip at $\Delta_{d}=0$. The $\sim10$-MHz shifts of the EIT peaks for $\Delta_d /(2 \pi)= \pm 20~$MHz from $\Delta_{c}=0$ are also reproduced.

\subsubsection{Analytical model and comparison with numerical model}
\label{subsubsec:amod}

To understand the results, it helps to first consider an analytical model for the case of weak probe ($\Omega_p \lesssim 2 \pi \times 1$~MHz), large dressing and coupler Rabi frequencies, and no atomic decay. In this case, the strongly-coupled three-level subspace $\{ \vert e \rangle, \vert d \rangle, \vert r_1 \rangle \}$ has, in a two-frequency dressed-atom picture and $\Delta_p=0$, a Hamiltonian given by
\begin{equation}
H_{\mathrm{sub}} (v) = \hbar
\left(
  \begin{array}{ccc}
        v k_1 & \Omega_{d}/2 & 0 \\
    \Omega_{d}/2 & - \Delta_d + v k_2  & \Omega_{c}/2 \\
    0 & \Omega_{c}/2 & - \Delta_d - \Delta_c + v k_3
  \end{array}
\right) \quad,
\end{equation}
with $k_1 = k_p$, $k_2=k_p-k_d$ and $k_3=k_p-k_d \mp k_c$ for the $(+, -, \mp)$ configurations.
Since the microwave is off here, $\vert r_2 \rangle$ is not coupled.

Absorption on the probe transition occurs for eigenstates with eigenvalue $s=0$, {\sl{i.e.}} we solve

\begin{equation}
\label{eq:roots}
H_{\mathrm{sub}} (v) \,
                \left(
                  \begin{array}{c}
                    c_e \\
                    c_d \\
                    c_r \\
                  \end{array}
                \right) = s
                \left(
                  \begin{array}{c}
                    c_e \\
                    c_d \\
                    c_r \\
                  \end{array}
                \right) =
                \left(
                  \begin{array}{c}
                    0 \\
                    0 \\
                    0 \\
                  \end{array}
                \right)
                \quad,
\end{equation}

where the $c_i$ are the coefficients of a normalized eigenstate. Solving det$(H_{\mathrm{sub}} (v))=0$ amounts to finding the roots of a third-order polynomial in $v$, which has real solutions $v_l$ with a counter $l$ ranging from 1 to up to 3. The state coefficients $c_{i,l}$ then follow for each of the real roots, $v_l$. The strength of the probe absorption of atoms traveling at velocity $v_l$ is proportional to $\vert c_{e,l} \vert^2$, and the net absorption summed over all roots is approximately proportional to $\sum_l P(v_l) \vert c_{e,l} \vert^2$. Here, we obtain the roots $v_l$ as a function of $\Delta_c$, for selected values of $\Delta_d$, and plot them on the $(\Delta_c, v)$-plane. Using the Rabi frequencies $\Omega_d$ and $\Omega_c$ listed in Fig.~\ref{fig:2}, we plot the roots as crosses in Figs.~3~(a) and~(b); symbol diameter is proportional to $\vert c_{e,l} \vert$.

Our numerical model for the absorption, outlined in Sec.~\ref{sec:model}, is more accurate because it accounts for the level decays and probe saturation. Numerically solving Eqs.~(1-5) yields $[d\alpha/dv] (\Delta_c, v)$, which we also plot in Figs.~3~(a-c) on a color map using the $\Omega_p$, $\Omega_d$ and $\Omega_c$ listed in Fig.~\ref{fig:2}. The absorption coefficient $\alpha(\Delta_c)$ is obtained by integrating $d\alpha/dv$ over $v$. The resultant transmission spectra, plotted in Fig.~3~(d), are $T(\Delta_c) = \exp(-\alpha \, L)$ with cell length $L=7.5$~cm. In Figs.~3~(a-c) it is seen that the analytical roots $v_l$ and $\vert c_{e,l} \vert$-values from Eqs.~(6-7) and the numerical solutions from Eqs.~(1-5) present the same picture as to which velocity classes in the atomic vapor cause what degree of absorption. Thereby, the analytical model from Eqs.~(6-7) is useful because it lends itself to elucidate the underlying physics. On the other hand, the numerical solutions from Eqs.~(1-5) are required to quantitatively model the experimentally measured spectra. In the following we will use both the analytical model and the numerical solutions to interpret the various observed spectra.

\subsubsection{Interpretation of results}

\begin{figure*}[t]
\centering
\includegraphics[width=0.9\textwidth]{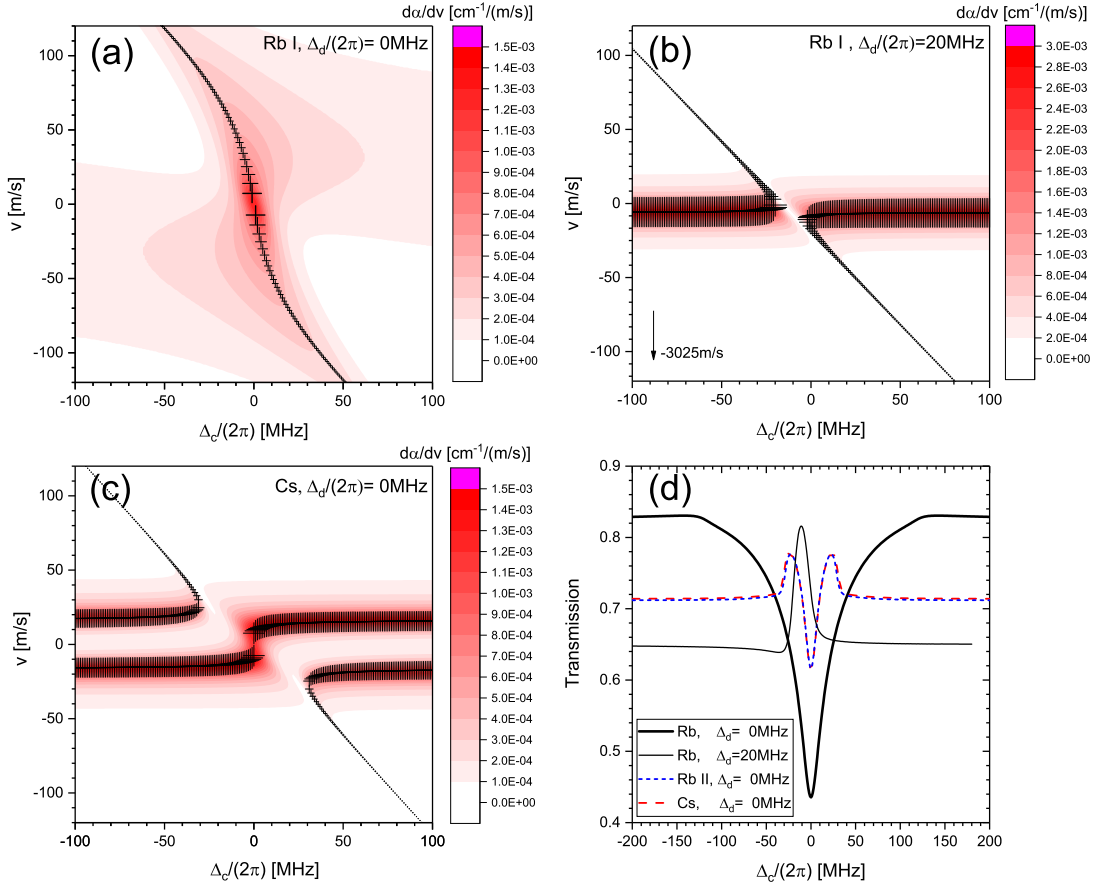}
\caption{(a-c) Velocity-specific absorption-coefficient maps, $[d\alpha/dv] (\Delta_c, v)$, vs coupler detuning and velocity, displayed on linear color maps, for $(+, -, -)$ configuration and the indicated values of the dressing-beam detuning, $\Delta_d$, calculated with the complete numerical model from Sec.~\ref{sec:model}. The probe detuning $\Delta_p=0$, and Rabi frequencies are as in Fig.~2~(b). The crosses and their diameters show positions and absorption strengths of dressed states derived from the analytical model explained in Sec.~\ref{subsubsec:amod}, where atomic decay is neglected and the probe is assumed to be weak. Panels (a) and (b) correspond to EIA and EIT, respectively, for parameters as used in our rubidium experiment and cell temperature 300~K. For comparison, in (c) we show an absorption map and dressed-state positions and absorptions for
EIA on the cesium cascade with wavelengths $\lambda_p=852$~nm, $\lambda_d=1470$~nm and $\lambda_c=790$~nm. To allow for a comparison, atom density, velocity distribution $P(v)$ and Rabi frequencies are the same in (a-c). Panel (d) shows probe transmissions obtained from panels (a-c), and for the Rb cascade with wavelengths $\lambda_p=780$~nm, $\lambda_d=1366$~nm and $\lambda_c=740$~nm (labeled Rb II).} 
\label{fig:3}
\end{figure*}

We first discuss the case of EIA. In the middle curves in Fig.~2, the parameters are $\Delta_d=0$, $k_2 < 0$ and $\Omega_c/\Omega_d < \sqrt{-k_3/k_1} = 1.266$. For this case it is found that Eq.~(\ref{eq:roots}) has only one real root, $v_1$, at any $\Delta_c$. Fig.~\ref{fig:3}~(a) shows $[d\alpha/dv] (\Delta_c, v)$ and the root $v_1$ of Eq.~(\ref{eq:roots}) for this case. The root closely follows the ``ridge line'' of large  $d\alpha/dv$ obtained in the exact calculation (dark-red regions on the color map). Also, $\vert c_{e,l} \vert$, indicated by symbol size, presents a good qualitative measure for the magnitude of $d\alpha/dv$ along the ridge line.
Integrating the $d\alpha/dv$-data in Fig.~\ref{fig:3}~(a) over $v$ yields the thick solid curve in Fig.~\ref{fig:3}~(d). In view of Fig.~\ref{fig:3}~(a), it is apparent that EIA becomes particularly strong when the derivative of the root $dv_1/d\Delta_c$ at $\Delta_c=0$ becomes large. In this case, absorption from a wide range of velocity classes in the vapor cell is accumulated at $\Delta_c \approx 0$, leading to particularly strong EIA. As previously discussed in~\cite{carr_three-photon_2012}, the EIA feature is deepest and narrowest when $\Omega_c/\Omega_d = \sqrt{-k_3/k_1} = 1.266$. This condition is equivalent to $[dv_1/d\Delta_c] (\Delta_c = 0)$ $\rightarrow \infty$.

For the Rb cascade studied in our experiment, EIA is enhanced even more because $k_2 \approx 0$ for this cascade. For $k_2 \approx 0$  the region of large $d\alpha/dv$ at $\Delta_c \approx 0$ extends over a particularly wide range in velocity (see Fig.~\ref{fig:3}~(a)), leading to a large integral $[ \int [d \alpha/dv ] dv] (\Delta_c)$, as evident in the thick solid curve in Fig.~\ref{fig:3}~(d). In contrast, for large and positive $k_2$, the case of \cite{carr_three-photon_2012}, one finds that Eq.~(\ref{eq:roots}) has three roots in most $\Delta_c$-domains and that the region of large $d\alpha/dv$ at $\Delta_c = 0$ is limited to $\vert v \vert \lesssim \Omega_d /(2 \sqrt{k_1 \, k_2})$. This is visualized in Fig.~\ref{fig:3}~(c), which is for the Cs transitions chosen in~\cite{carr_three-photon_2012}. There, for $\Omega_d = 2 \pi \times 25$~MHz the region of large $d\alpha/dv$ is capped at $\vert v \vert \lesssim \Omega_d /(2 \sqrt{k_1 \, k_2}) = 16.5$~m/s, leading to a comparatively small EIA effect at $\Delta_c \approx 0$. The three transmission curves plotted in Fig.~\ref{fig:3}~(d) for $\Delta_d = 0$ demonstrate that the Rb case with $\lambda_p=780$~nm and $\lambda_d=776$~nm has, indeed, by far the strongest EIA.

For comparison, in Fig.~\ref{fig:3}~(d) we additionally show the EIA curve for the 5$S_{1/2} \leftrightarrow5P_{3/2}\leftrightarrow 6S_{1/2}\leftrightarrow nP_J$ cascade in Rb, which has wavelengths $\lambda_p=780$~nm, $\lambda_d=1366$~nm, $\lambda_c=740$~nm, and $k_2 \gg 0$. It is seen that this case exhibits relatively weak EIA, similar to that of the Cs cascade studied in~\cite{carr_three-photon_2012}.

We now briefly discuss the case of EIT for off-resonant dressing beam and
then consider the context between EIA and EIT.
In Fig.~\ref{fig:3}~(b) we show the roots $v_l$ of Eq.~(\ref{eq:roots}) and the exact solution for $d\alpha/dv$ for the case $\Delta_d = 2 \pi \times 20$~MHz (top curves in Fig.~2). Except for $\Delta_c \sim 0$, there are three roots, with the root near $v=-3025$~m/s having zero absorption. The other two roots form a comparatively narrow anti-crossing that results in correspondingly narrow EIT signals (Fig.~2 top and bottom panels and blue curve in Fig.~\ref{fig:3}~(d)).

The difference in behavior seen at zero vs substantially non-zero $\Delta_d$ (strong, wider EIA versus somewhat less strong, narrower EIT) corresponds to different limits of the dressed states formed by the dressing transition. The
velocity roots that correspond to the two dressed states at large $\Delta_c$ (or, with the coupler turned off) are given by $v_\pm = \frac{\Delta_d}{2 k_2} \pm \sqrt{(\frac{\Delta_d}{2 k_2})^2 + \frac{\Omega_d^2}{4 k_1 k_2}}$.
For $\Delta_d=0$ and $k_2>0$, resonant coupling results in a pair of symmetric and anti-symmetric Autler-Townes(AT)-split states at $v_\pm = \pm \frac{\Omega_d}{2 \sqrt{k_1 k_2}}$ that both have $50\%$ probability in $\vert e \rangle$, leading to two equally strong horizontal absorption bands, as seen in Fig.~\ref{fig:3}~(c).
The width of the EIA feature near $\Delta_c=0$ scales with $\Omega_c$ (see Fig.~\ref{fig:3}~(c) and the cases of $\Delta_d =0$ in Fig.~\ref{fig:3}~(d)). Note that in Fig.~\ref{fig:3}~(a) the horizontal absorption bands are absent because $k_2 < 0$.
On the other hand, for large $\Delta_d$ the two AT states are highly asymmetric. In that case, the dressing and coupler beams drive a two-photon transition that has intermediate detuning $\Delta_d$ from the $\vert d \rangle$-level and two-photon Rabi frequency $\Omega_d \, \Omega_c / (2 \Delta_d)$. This leads to narrow EIT lines at large $\Delta_d$, with widths on the order of $\Omega_d \, \Omega_c / (2 \Delta_d)$ (see Fig.~\ref{fig:3}~(b) and the cases of $\Delta_d = 2 \pi \times 20$-MHz in Fig.~\ref{fig:3}~(d)).

\subsection{($+, -, +$) configuration and comparison}

\begin{figure}[htb]
\centering
\includegraphics[width=0.48\textwidth]{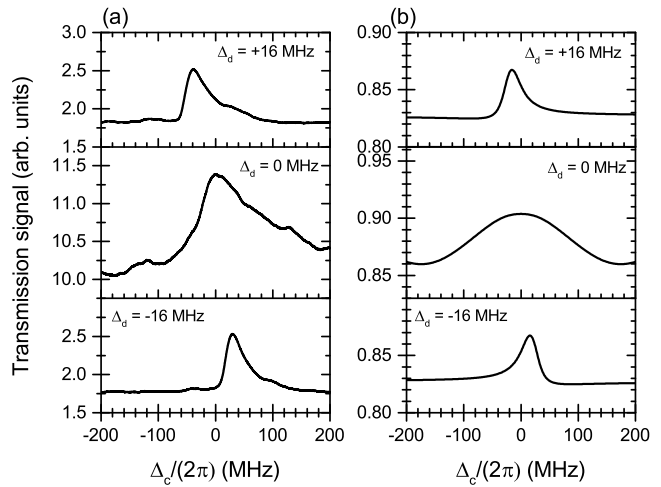}
\caption{(a) Experimental probe transmission signal vs $\Delta_c$ for the indicated values $\Delta_d$, for laser beam directions in the ($+, -, +$) configuration. The transmission axes are on the same arbitrary scale.
EIT is observed for all cases with $\Delta_d /(2 \pi) = -16, 0, +16$~ MHz. (b) Probe transmissions calculated with the model in Sec.~\ref{sec:model} for cell temperature of 300~K, cell length $L=7.5$~cm, and $\Omega_p =  2 \pi \times 15$~MHz, $\Omega_d = 2 \pi \times 30$~MHz, and $\Omega_c=   2 \pi \times 18$~MHz. }
\label{fig:4}
\end{figure}

Figure~\ref{fig:4}~(a) shows experimental results when the laser beams are in ($+, -, +$) configuration. We obtain strong EIT signals when $\Delta_d$ is close to zero and weak EIT signals when $\Delta_d = \pm 2 \pi \times 16$~MHz. The results show reasonable agreement with the simulation in Fig.~\ref{fig:4}~(b). In both experiment and simulation it is seen that the $\Delta_d=0$-case exhibits wide, massive EIT over the entire range, with a broad peak around $\Delta_c=0$, and narrower, less high and asymmetric EIT peaks in the detuned-$\Delta_d$ cases. In the detuned-$\Delta_d$ cases, the EIT peaks are shifted from $\Delta_c=0$, both in experimental and simulated results. Measured and simulated spectra deviate from each other in that in all cases the experimental EIT peaks are lopsided to the right. This may be attributable to slight elliptical polarizations and optical pumping of the atoms within the interaction volume. Such effects are not covered by our model because it neglects magnetic substructure.

\begin{figure*}[t]
\centering
\includegraphics[width=0.9\textwidth]{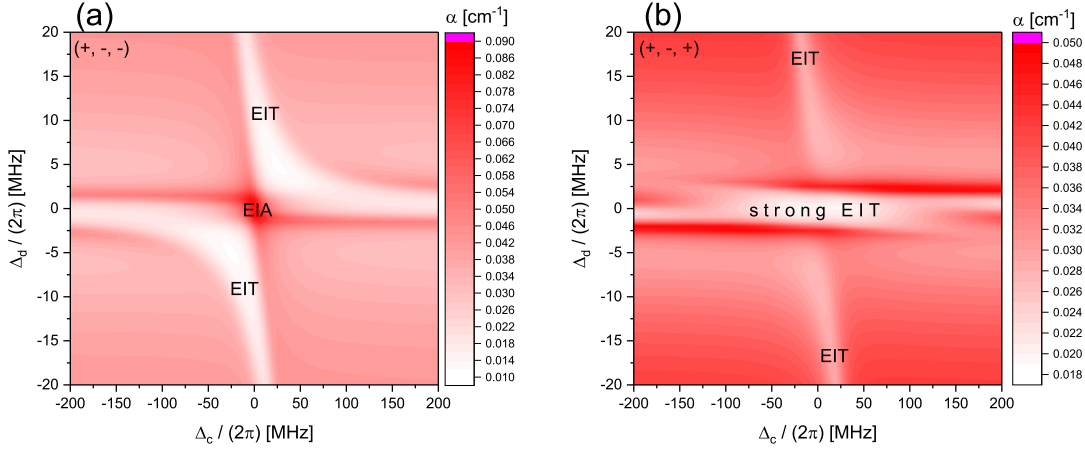}
\caption{Absorption coefficient, calculated with the model in Sec.~\ref{sec:model}, for Rb three-photon EIT/EIA for the level scheme shown in Fig.~\ref{fig:1} (without microwave) vs coupler (horizontal) and dressing detuning (vertical axis) for the ($+, -, -$) (panel a) and the ($+, -, +$)  (panel b) beam-direction configurations. To allow for a comparison, in both (a) and (b) we have used the same Rabi frequencies, $\Omega_p =2 \pi \times 10$~MHz, $\Omega_d=2 \pi \times 25$~MHz  and $\Omega_c=2 \pi \times 18$~MHz.
}
\label{fig:5}
\end{figure*}

In Fig.~\ref{fig:5} we present an overview calculation to compare the EIT and EIA effects between the ($+, -, \mp$) configurations. The absorption coefficient is plotted vs coupler-laser scan, $\Delta_c$ (horizontal axis), for a range of dressing-beam detunings, $\Delta_d$ (vertical axis). All phenomena explained above are reproduced. Noting the difference in color-scale range, it is reaffirmed that the ($+, -, +$) case generally exhibits much stronger EIT than the ($+, -, -$)-case, across all frequency detunings.

In terms of linewidth and signal height/ depth, the Rydberg-EIT/EIA features in the ($+, -, \mp$) configurations can be ranked in usability for Rydberg-state spectroscopy in the following descending order:

\noindent (1) ($+, -, -$)-EIT \\
(2) ($+, -, -$)-EIA \\
(3) ($+, -, +$)-EIT for $\Delta_d \ne 0$ \\
(4) ($+, -, +$)-EIT for $\Delta_d = 0$,

In the next section we calibrate a 100-GHz transmission system using EIT and EIA in the ($+, -, -$) configuration, the measurement methods that rank the highest in our comparison.

\section{Microwave measurements}
\label{sec:onephoton}

Rydberg spectroscopy presents an excellent tool for microwave~\citep{Holloway2014} and sub-THz~\cite{AndersonGSMM2018} metrology. Using three-photon Rydberg-EIT/EIA with red and infrared laser diodes may present an advantage over the more widely used two-photon Rydberg-EIT due to reduced cost and the fact that red and infrared light may cause less photoelectric effect and ionization within the vapor cells, potentially reducing the effects of DC electric fields on the quality of the Rydberg spectra. Here we use microwave-field-induced AT splitting on the Rb $28F_{7/2}\leftrightarrow 29D_{5/2}$ resonance to measure a microwave field.

\begin{figure}[htb]
\centering
\includegraphics[width=0.5\textwidth]{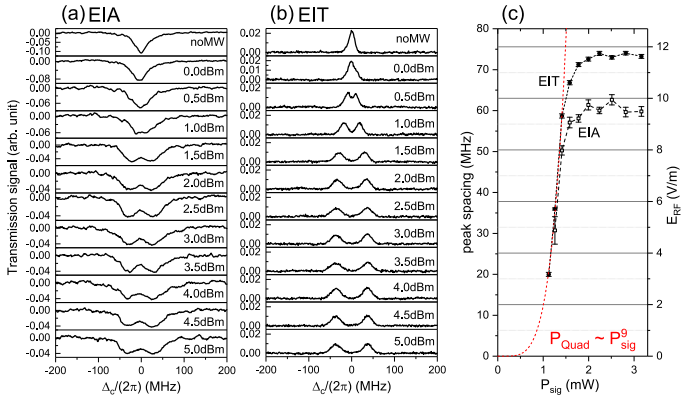}
\caption{
(a) A series of EIA spectra in ($+,-,-$) configuration with $\Delta_d=0$ and a microwave field of 100.633~GHz driving the $28F_{7/2} \leftrightarrow 29D_{5/2}$  transition. The probe field is horizontally and all other fields are vertically polarized. The spectra are labeled by the power $P_{sig}$ the signal generator supplies to the frequency quadrupler. (b) Same for EIT in ($+,-,-$) configuration with $\Delta_d=- 2 \pi \times 80$~MHz.
(c) Autler-Townes splittings (left axis) from panels (a) and (b) and derived microwave electric fields (right axis) vs $P_{sig}$. An allometric fit to the EIT data shows that in the non-saturated regime the quadrupler power scales with the ninth power of $P_{sig}$.}
\label{fig:6}
\end{figure}

The $28F_{7/2} \leftrightarrow 29D_{5/2}$  transition has a large radial electric-dipole moment matrix element of $1047$~ea$_0$ and an average angular matrix element for the relevant $\pi$-polarized transitions ($m=1/2$ and $3/2$) of $0.47$. The effective electric-dipole moment $\bar{d}_z$ is the product of the radial and averaged angular matrix elements. The microwave field, $E_{RF}$, follows from

\begin{equation}
\label{eq:at}
E_{RF} \approx \frac{\hbar \Delta_{AT}}{\bar{d}_z} \quad ,
\end{equation}

with AT splitting $\Delta_{AT}$ and $\bar{d}_z=492$~ea$_0$. The large value of $\bar{d}_z$ makes the measurement method very sensitive to the microwave field. We set the microwave generator (Keysight N5183A MXG) at a frequency of 25.1582~GHz. The field is frequency-quadrupled using an active frequency multiplier (Norden N14-4680) to reach the microwave frequency of 100.633~GHz, which is on resonance with the $28F_{7/2} \rightarrow 29D_{5/2}$ transition. The AT splitting observed in three-photon Rydberg-EIT/EIA approximates the microwave Rabi frequency, $\Omega_{RF}$, which in turn reveals the microwave electric field according to Eq.~(\ref{eq:at})~\citep{sedlacek_microwave_2012, holloway_electric_2017, Kuebler2019}.

In Fig.~\ref{fig:6} we present measurements for both EIA (left) and EIT (middle) in the ($+, -, -$) configuration, as well as the derived AT splittings (right), for the indicated power levels of the $25$-GHz signal generator. It is evident that the EIT signals have a narrower linewidth, allowing one to resolve the AT peaks at a lower microwave field than in the EIA case. In the present case, the AT-splitting data allow us to perform an absolute calibration of the 100-GHz microwave electric field at the location of the vapor cell relative to the utilized microwave horn (Chendu LB-10-15). The AT splittings (left axis in Fig.~\ref{fig:6}~(c)) and Eq.~(\ref{eq:at}) yield the RF electric field (right axis). The data in Fig.~\ref{fig:6}~(c) show that below saturation the quadrupler power scales as the ninth power of the signal-generator power, highlighting the fact that the quadrupler is a highly nonlinear device.

The saturation RF electric field of the quadrupler at the atom location is about 11~V/m, as seen in Fig.~\ref{fig:6}~(c). Using the standard gain from the horn manufacturer's data sheet, 18.7~dBi, and the chosen distance between the horn and the cell, 28~cm (which is in the far field), the maximum radiated power from the quadrupler is estimated at 2~mW. The quadrupler data sheet specifies 1~mW. The slight elevation of our power measurement may be due to constructive standing-wave interference of the 100-GHz field within the cell~\cite{holloway_sub-wavelength_2014,
Fan2014, Fan2015, Holloway2017TEMC}, which would increase the measured output power of the quadrupler, and/or to a conservative quadrupler specification ({\sl{i.e.}}, the quadrupler might perform slightly better than specified). The change in laser beam paths between the EIA and EIT sets of data shown in Fig.~\ref{fig:6} may have caused a slight variation in standing-wave effects, which could explain the difference in saturation electric fields between the EIT and EIA measurements of the RF electric field.

\begin{figure}[tb!]
\centering
\includegraphics[width=0.5\textwidth]{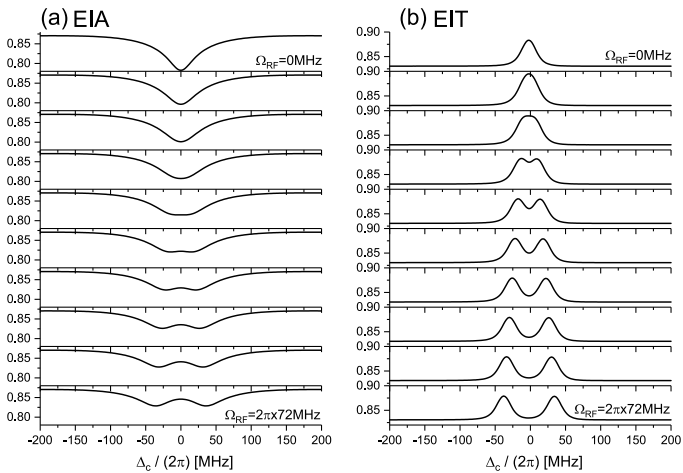}
\caption{Calculated EIA (a) and EIT (b) transmission spectra with the RF field turned on, for $\Delta_d =0$ and
$\Delta_d=- 2 \pi \times 80$~MHz, respectively. The laser beams are in ($+,-,-$) configuration, and the optical Rabi frequencies are $\Omega_p =2 \pi \times 20$~MHz, $\Omega_d=2 \pi \times 25$~MHz  and $\Omega_c=2 \pi \times 18$~MHz. The RF Rabi frequency ranges from 0 (top) to $2 \pi \times 72$~MHz (top), varied in equidistant steps.}
\label{fig:7}
\end{figure}

Finally, we have modeled the RF spectra in Fig.~\ref{fig:6} along the lines of Eqs.~(1-5). The results, shown in Fig.~\ref{fig:7}, are in good agreement with the measured spectra. It is, in particular, confirmed that the EIT signal at large dressing-beam detuning $\Delta_d$ allows one to resolve smaller AT spittings than the EIA signal, due to the smaller width of the EIT peaks.

\section{Conclusion}

We have performed a comprehensive experimental and theoretical study of three-photon Rydberg-EIA/EIT in an atomic vapor cell.
Physical interpretations have been provided that elucidate the underlying physics.
Fig.~\ref{fig:6} demonstrates that three-photon Rydberg-EIT, with low-cost all-infrared laser diode systems, may be valuable for absolute calibration of microwave frequency instrumentation. This could be particularly useful in the sub-THz and THz frequency regimes, where detectors can be inaccurate or may be unavailable. In future work it is desirable to account for optical-pumping effects, as well as for line splittings in complex RF spectra (for instance, spectra obtained in stronger RF fields or with more highly-excited Rydberg levels). The large Hilbert spaces in such extended Rydberg-EIT/EIA systems can be modeled efficiently using quantum Monte Carlo wavefunction methods~\cite{Zhang2018, Xue2019} and Floquet methods~\cite{Anderson2014, Anderson2016}, respectively.

\section{Acknowledgements}
This work was supported by the NSF (Grants No. PHY-1806809 and PHY-1707377). DAA acknowledges support by Rydberg Technologies Inc.

Present address:\\
$^{\dag}$~Universit\"at Heidelberg, Heidelberg 69120, Germany\\
$^{\dag\dag}$~National Institute of Standards and Technology, Boulder, Colorado  80305, USA\\
$^{\dag\dag\dag}$~Michigan State University, East Lansing, Michigan 48824, USA\\
$^{\dag\dag\dag\dag}$~Extreme Light Infrastructure (ELI-NP), Str. Reactorului No. 30, 077125 Bucharest-M\u{a}gurele, Romania\\

\bibstyle{apsrev4-1}
%

\end{document}